\documentstyle{article}
\def\be{\begin{equation}}
\def\ee{\end{equation}}
\def\bea{\begin{eqnarray}}
\def\eea{\end{eqnarray}}

\def\l{\label}

\begin{document}

\hfill{}

\vskip 3
 \baselineskip
 \noindent

 {\Large\bf Spin Zero Quantum Relativistic  Particles in Einstein Universe}

\vskip \baselineskip

M. R. Setare {\footnote {E-mail: rzakord@ipm.ir}, Khaled
Saaidi{\footnote {E-mail: KSaaidi@ipm.ir} \vskip\baselineskip

 {\small Institute for Studies in Theoretical Physics and Mathematics,
 P.O.Box, 19395-5531, Tehran, Iran}\\

 {\small Department of Science,  University of
 Kurdistan, Pasdaran Ave., Sanandaj, Iran }

\vskip 2
\baselineskip



\vskip 2\baselineskip

\begin{abstract}
In this letter we have considered the eigenvalues and
eigenfunctions of relativistic massless scalar particle which
conformally coupled to the background of Einstein universe. We
found the eigenvalues and eigenfunctions exactly.
\end{abstract}

\newpage
\section{Introduction}
The one of the most interesting problem in theoretical physics is
the connection between the quantum mechanics and gravity. Since
the birth of quantum mechanics there has been many work in this
topic \cite{lan}-\cite{vak}. Chandrasekhar, studied the Dirac
equation in a Kerr space-time background. He separated the Dirac
equation into radial and angular parts\cite{6}. The Page extended
this work to the Dirac equation for an electron around a
Kerr-Newman back hole back ground, in this case also the Dirac
equation is separated into decoupled ordinary differential
equation\cite{10}. The separation of variables  for the massive
complex Dirac equation in the gravitational back ground of the
Dyon black hole has been done in \cite{11}. Also, in \cite{9} the
radial part of Dirac equation in a Schwarzschild geometry has been
solved by using WKB approximation method. Moreover above
theoretical works several experiments have been performed to test
theoretical predications, for example the authors of \cite{4} have
been used a neutron interferometer to observe the quantum
mechanical phase shift of neutrons caused by their interaction
with the gravitational field of Earth. Furthermore, Nesvizhevsky
et al \cite{N} have been measured the quantum energy levels of
neutron in the Earth's gravitational field. In this letter we
shall consider the eigenvalues and eigenfunctions of relativistic
massless scalar particle which conformally coupled to the
background of Einstein universe. We will find the eigenvalues and
eigenfunctions exactly.

\section{Massless conformally coupled spin zero quantum relativistic  particles}

The simplest cases in which the analysis of quantum behavior of
particles can be done is the massless spin zero particles in the
static spacetime. Here we consider the Einstein universe, which
scalar particle conformally coupled it. The line element for, this
space is given by

\be\l{1} ds^2 = dt^2 - a^2{d\chi^2 + \sin(\chi)^2(d\theta^2 +
\sin^2\theta d\varphi^2)}, \ee

in this case the scalar curvature is as following \be\l{2} R = {6
\over a^2 }. \ee

The equation of motion for massless scalar particles is given by
\be\l{3} (\Box + \xi R)u(x) = 0, \ee where $\Box$    is given by
\be\l{4} \Box = (-g)^{1 \over 2} \partial_\mu[(-g)^{1\over 2}
g^{\mu\nu} \partial_\nu]. \ee

The modes $u(x)$ are separated as \be\l{5} u_k(x) = {1 \over
a}y_k(x) \chi_k(t), \ee

with $x = (\chi, \theta, \varphi )$      and $y_k(x)$ is a
solution of \be\l{6} \Delta^3y_k(x) = -(k^2-1)y_k(x),\ee with this
separation $\chi_k$       satisfies following equation \be\l{7}
{d^2\chi_k \over d\eta^2} + {\biggl [}{ k^2 + c(\eta)[\xi -
\xi(4)]R(\eta)}{\biggr ]}\chi_k = 0, \ee

with  $\xi(4) = {1 \over 6}$, also  $c(\eta )= a^2(t)$, which the
conformal time parameter  $\eta$   given by \be\l{8} \eta =
\int^ta^{-1}(t') dt'.\ee

Using the scalar curvature Eq(2), the normalized solution of Eq(1)
can be written as \be\l{9} \chi_k(\eta) = (2\omega_k)^{-1/2}
e^{-i\omega_k \eta }, \ee where \be\l{10} \omega_k^2 = k^2 + (6\xi
-1). \ee

For conformally coupled case  $\xi= {1 \over 6}$  we have
\be\l{11} \omega_k^2 = k^2, \hspace{2cm}k=1,2,...\ee Which
$\omega_{k}=k$  are eigenvalues of massless scalar particles in
our interest background. Therefore the eigenvalues are equally
spacing. The eigenfunctions $y_k$ of the three-dimensional
Laplacian are \cite{{par},{da}}

\be\l{12} y_k(x)= \Pi_{Kj}(\chi) Y^m_j(\theta , \varphi),
\hspace{3cm} K = (k, j, M)
 \ee

where \be\l{13} M= -j, -j+1, \ldots, j ; \hspace{1cm} j = 0, 1,
\ldots, k-1 \hspace{1cm} k= 1, 2, \ldots \ee

The $Y^M_j$    are spherical harmonics. The functions
   ${\Pi^{(+)}_{Kj}}(\chi)$ are defined by \cite{lif}
   \be
   \Pi^{(+)}_{Kj}(\chi) = {\biggl \{}{1 \over 2}\pi(ik)^2  [(ik)^2 + 1^{2}]\ldots [(ik)^2 +j^2]{\biggr \}}^{-1 \over
   2} \sinh^j(i\chi){\biggl(}{d \over d\cosh(i\chi)}{\biggr )}^{1 + j} \cos(k\chi).   \ee

\end{document}